\documentclass[lettersize,journal]{IEEEtran}
\usepackage[T1]{fontenc}
\usepackage{amsmath,amsfonts}
\usepackage{algorithmic}
\usepackage{algorithm}
\usepackage{array}
\usepackage[caption=false,font=normalsize,labelfont=sf,textfont=sf]{subfig}
\usepackage{textcomp}
\usepackage{stfloats}
\usepackage{url}
\usepackage{verbatim}
\usepackage{graphicx}
\usepackage{cite}
\begin{document}
	
	\title{Assessment of sub-sampling schemes for compressive nano-FTIR imaging}
	
	\author{
    \IEEEauthorblockN{Selma Metzner\IEEEauthorrefmark{1}, Bernd Kästner\IEEEauthorrefmark{1}, Manuel Marschall\IEEEauthorrefmark{1}, Gerd Wübbeler\IEEEauthorrefmark{1}, Stefan Wundrack\IEEEauthorrefmark{1}, Andrey Bakin\IEEEauthorrefmark{2}, Arne Hoehl\IEEEauthorrefmark{1}, Eckart Rühl\IEEEauthorrefmark{3} and Clemens Elster\IEEEauthorrefmark{1}\\}
    \IEEEauthorblockA{\IEEEauthorrefmark{1}Physikalisch-Technische Bundesanstalt, Braunschweig and Berlin, Germany\\}
    \IEEEauthorblockA{\IEEEauthorrefmark{2}Institut für Halbleitertechnik, Technische Universität Braunschweig, Hans-Sommer Straße 66, 38106 Braunschweig, Germany\\}
    \IEEEauthorblockA{\IEEEauthorrefmark{3}Physikalische Chemie, Freie Universität Berlin, Arnimallee 22, 14195 Berlin, Germany}}
	
	
	
	\maketitle
	
	\begin{abstract}
		Nano-FTIR imaging is a powerful scanning-based technique at nanometer spatial resolution which combines Fourier transform infrared spectroscopy (FTIR) and scattering-type scanning near-field optical microscopy (s-SNOM). However, recording large spatial areas with nano-FTIR is limited by long measurement times due to its sequential data acquisition. Several mathematical approaches have been proposed to tackle this problem. All of them have in common that only a small fraction of randomly chosen measurements is required. 
		However, choosing the fraction of measurements in a random fashion poses practical challenges for scanning procedures and does not lead to time savings as large as desired. We consider different, practically relevant sub-sampling schemes assuring a faster acquisition. It is demonstrated that results for almost all considered sub-sampling schemes, namely original Lissajous, triangle Lissajous, and random reflection sub-sampling, at a sub-sampling rate of 10\%, are comparable to results when using a random sub-sampling of 10\%. This implies that random sub-sampling is not required for efficient data acquisition.
	\end{abstract} 
	

\section{Introduction}

Infrared (IR) spectroscopy is a non-destructive technique for material characterization ranging from analytical chemistry~\cite{salzer2009,morsch2020}, materials sciences \cite{ruggeri2015,chae2015}, life sciences \cite{pilling2016,paluszkiewicz2017differentiation}  to microelectronics~\cite{lau1999}. While the spatial resolution of conventional IR
microscopy is limited to several micrometers due to the diffraction limit  \cite{born2013principles}, this limitation can be overcome by modern scanning probe microscopy-based methods in applying, for instance, near-field based techniques. Examples include scattering-type scanning near-field optical microscopy (s-SNOM) \cite{knoll1999near,keilmann2009near}, atomic force microscopy-based IR spectroscopy (AFM-IR) \cite{dazzi2005local,dazzi2017afm}, photoinduced force microscopy (PiFM) \cite{rajapaksa2010image}, tip-enhanced photoluminescence spectroscopy (TEPL) \cite{wang2021ambient}, and tip-enhanced Raman spectroscopy (TERS) \cite{stockle2000nanoscale}.  

Nano-FTIR \cite{huth2012nano,jones2012thermal,hermann2013near} is another scanning-based technique which combines Fourier transform infrared spectroscopy (FTIR) and s-SNOM. Here, FTIR spectroscopy uses a broadband IR source such as tunable lasers, thermal, or synchrotron radiation. The spatial resolution is enhanced in s-SNOM by adapting the principle of atomic force microscopy. Nano-FTIR enables the acquisition of the complete spectrum at each pixel of the scanned area with nanoscale spatial resolution. 

Usually, these scanning-based approaches take the spectra in a sequential process, resulting in long acquisition times for extended 2D arrays. These scans may take several hours to achieve nanometer spatial resolution over a substantial field of view in the micron size regime. In addition, long acquisition times may lead to sample and tip damage as well as drift artifacts. 

Several methods have been presented recently in order to overcome long acquisition times. In \cite{kastner2018compressed} a compressed sensing approach was introduced. Its potential benefit was demonstrated by reconstruction results achieved for nano-FTIR measurements when using only 11\% of the original data. A low-rank matrix reconstruction was developed in \cite{marschall2020} for FTIR measurements. An adequate sample recovery was achieved for a sub-sampling rate of 5\%. Both methods require only a small fraction of all measurements which could in principle reduce the experimental effort significantly. However, the fraction of measurements was chosen randomly. Practically, this implies that each single measurement position needs to be accessed individually, thereby also passing along idle routes and positioning times during which no measurement is carried out. 
Consequently, time savings are not as large as desired.

In this work, we introduce several, practically relevant sub-sampling schemes which sub-sample the data according to specific routes. The acquisition of these data avoids idle routes, is practically feasible and much faster than taking the same number of measurements at randomly chosen positions. Using the same low-rank matrix recovery method as in \cite{marschall2020} we compare the low-rank reconstructions of the fully sampled data set for different sub-sampling schemes with results obtained from random sub-sampling. The comparison is based on a data set for a sample corresponding to the edge of an ultrathin gallium film beneath epitaxial graphene grown on 6H-SiC.

The paper is organized as follows. In Section 2, the measurement process, data acquisition, and numerical low-rank approximation method are presented. Furthermore, four sub-sampling schemes will be described. Section 3 demonstrates results on different sub-sampling schemes. Finally, Section 4 gives some conclusions and an outlook to promising future research.

	\section{Methods}
	
	\subsection{Nano-FTIR}
	
	Nano-FTIR is a scanning probe technique which is based, on the one hand, on 
	Fourier-transform IR
	spectroscopy and, on the other hand, on atomic force microscopy. A sharp metallic tip which is brought close to the sample scans the sample surface and backscatters the incident IR radiation. The metallic tip behaves as an optical antenna and strongly restricts the incident optical field to the tip. The nearfield of the resulting nanoscale light source is then suitable for high-resolution imaging and spectroscopy. In order to separate the nearfield contribution from the farfield background the backscattered signal is demodulated at the second harmonic of the tip-modulation frequency. The interferogram is recorded by a Michelson interferometer \cite{hermann2013near,hermann2017enhancing} as a function of the optical path difference and the corresponding spectrum can be obtained from a Fourier transform of the interferogram.
	
	\subsection{Sample and data acquisition}
	As a sample, we chose the edge of an ultrathin gallium film beneath epitaxial graphene on 6H-SiC~\cite{wundrack2021liquid}. It results in a strongly resonant phononic spectrum around 920\,cm$^{-1}$ as well as a broad metallic spectrum, exhibiting in addition a transition region. Thus, we have a large spectral variation for benchmarking the proposed reconstruction algorithm. Details about fabrication and further sample properties can be found in the Appendix.
	
	Data acquisition from the Michelson interferometer results in an interferogram $I(z)$ with $N_z = 1024$ equidistant points ($\Delta z \approx 1.39~\mu$m). The interferogram is converted to a spectrum by Fourier transformation with a spectral resolution of $\Delta \Tilde{\nu} = 1/N_z \Delta z \approx 7.03~$cm$^{-1}$. Altogether, 961 interferograms were recorded on a rectangular area of $2\,\mu$m$ \times 2\,\mu$m using a Cartesian grid with 31$\times$31 spatial locations. The fully sampled nano-FTIR data set can thus be represented by a 3D data cube with dimensions of 31$\times$31$\times$1024, which was recorded within a total measurement time of 170 minutes.

	\subsection{Low-rank matrix reconstruction}
	
	The concept of low-rank matrices arises in many mathematical settings related to, e.g., modeling and data compression. A variety of applications ranges from signal processing \cite{weng2012low} and image restoration \cite{peng2014reweighted} to machine learning \cite{yao2015fast}. The recovery of a data matrix derived from incomplete observations poses a relevant example of tasks where low-rank techniques can be applied.
    In the present case, the data matrix is approximated by a matrix product with each factor having lower dimensionality, resulting in a recovery result of lower rank.
	The main idea of low-rank matrix reconstruction is that a low-rank approximation already captures the main characteristics of the data and less informative dimensions will be removed.
	In \cite{marschall2020}, a low-rank matrix reconstruction has been presented and successfully applied to sub-sampled FTIR data. In more detail, for observations $X$, the task is to find matrices $U$ and $V$ by minimizing $\sum_{i,j}|X_{i,j}-\sum_{k=1}^rU_{i,k}V_{j,k}|^2$. The so-called rank $r$ approximation $\hat{X}$ of $X$ is then given by $\hat{X}=UV^T$. However, the solution to this decomposition is non-unique. For this reason, and to stabilize the numerical reconstruction, an additional Tikhonov regularization \cite{engl1996regularization} has been utilized in \cite{marschall2020} which also accounts for the spatial smoothness of the focal images. We will use $r=20$ which was also used in~\cite{marschall2020}.
	
	As described above, the nano-FTIR data can be represented by a 3D data cube. The low-rank matrix reconstruction applies to a 2D object which will be handled in the following way: if $X$ denotes the matrix of measurements for the low-rank matrix reconstruction, then the rows of $X$ indicate the spatial position of the measurements and the columns of $X$ the interferometer position. Note that through a mapping from 3D to 2D (and vice versa) randomly chosen elements of the 2D matrix imply a specified selection of the 3D data cube (and vice versa).
	
	The resulting minimization problem is solved using an alternating algorithm which generates linear problems in each iteration. An implementation of the employed Python code has been made publicly available \cite{Software1}.

	\subsection{Sub-sampling schemes}
		\label{ch_sub}
	
	In the following, the considered sub-sampling schemes are presented.
	
	\begin{figure}[!t]
		\centering
		\subfloat[]{\includegraphics[width=.12\textwidth]{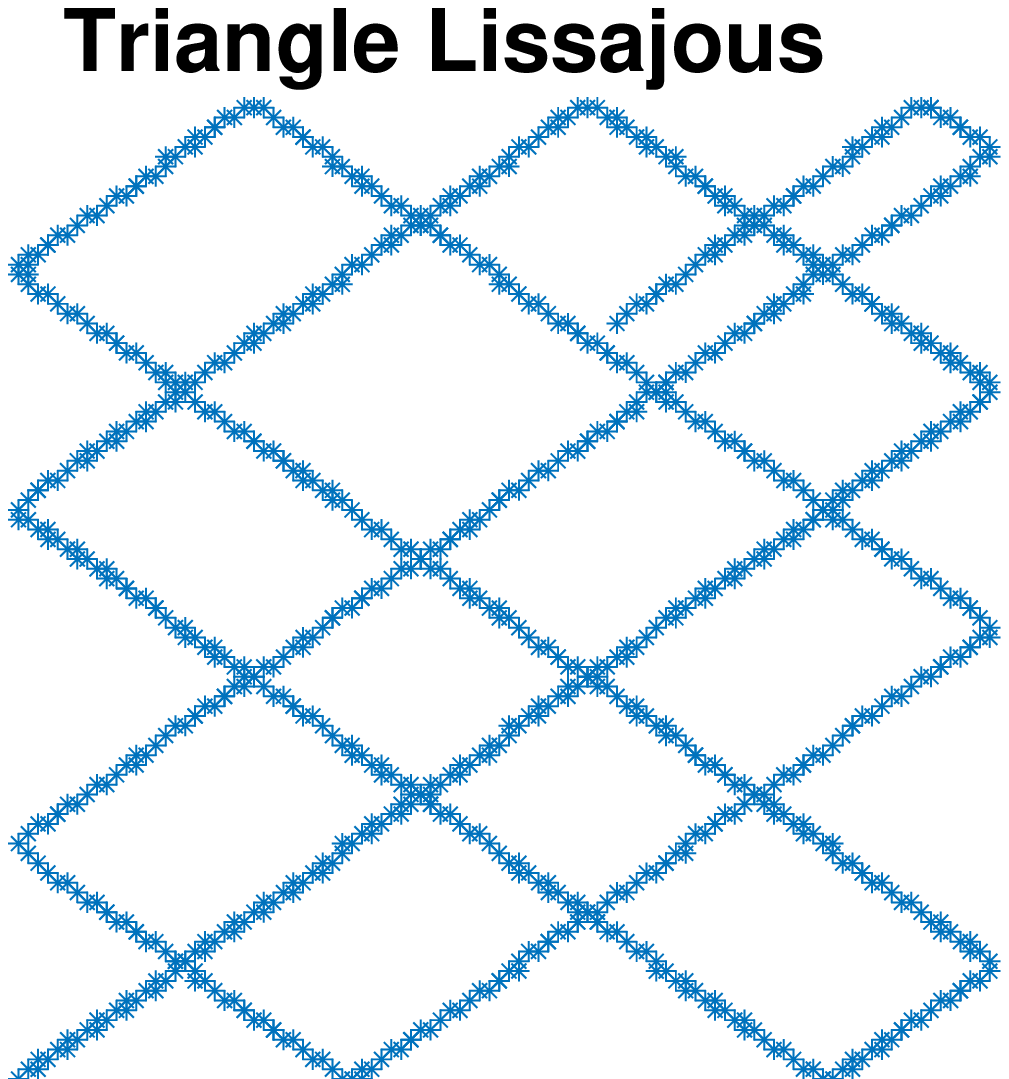}%
			\quad
			\label{fig_first_case}}
		\subfloat[]{\includegraphics[width=.12\textwidth]{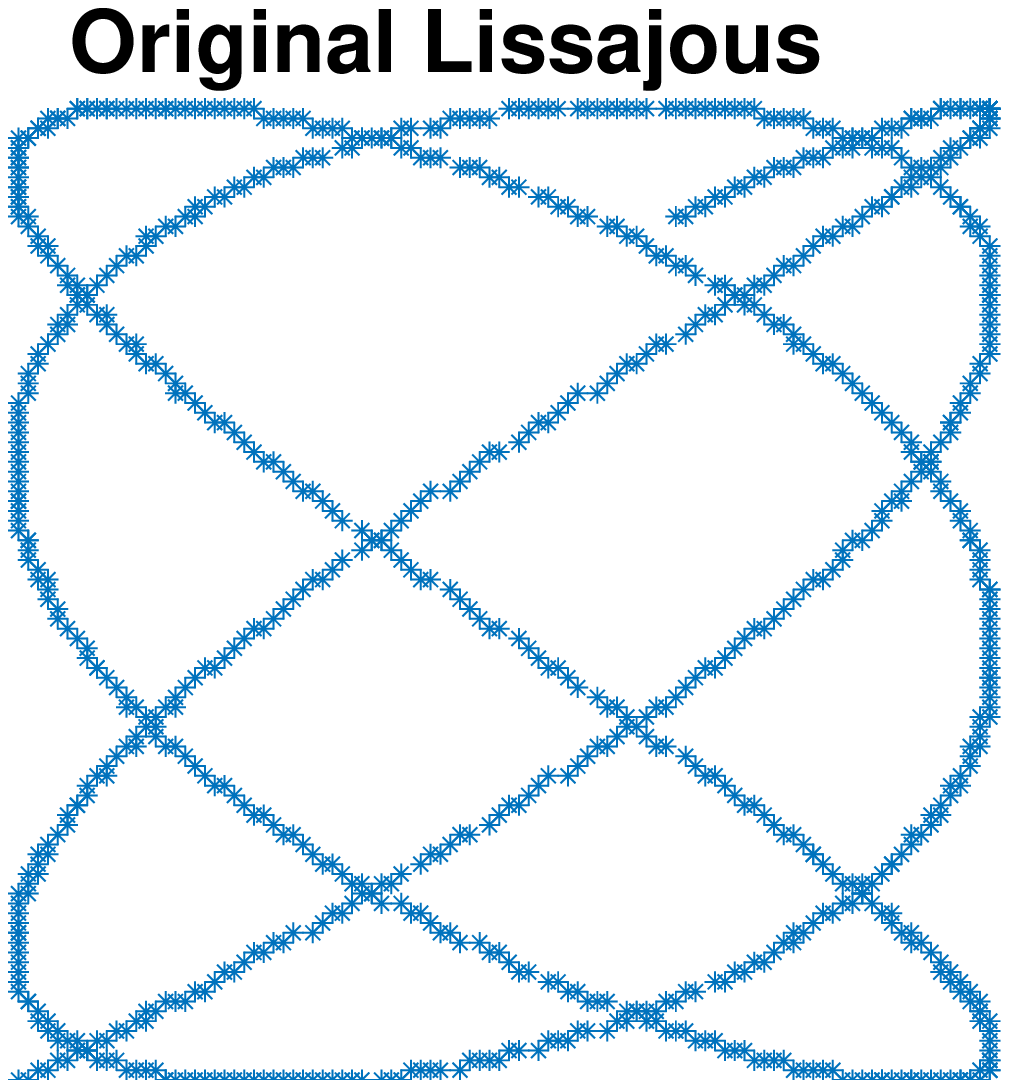}%
			\label{fig_second_case}}
		\quad
		\subfloat[]{\includegraphics[width=.12\textwidth]{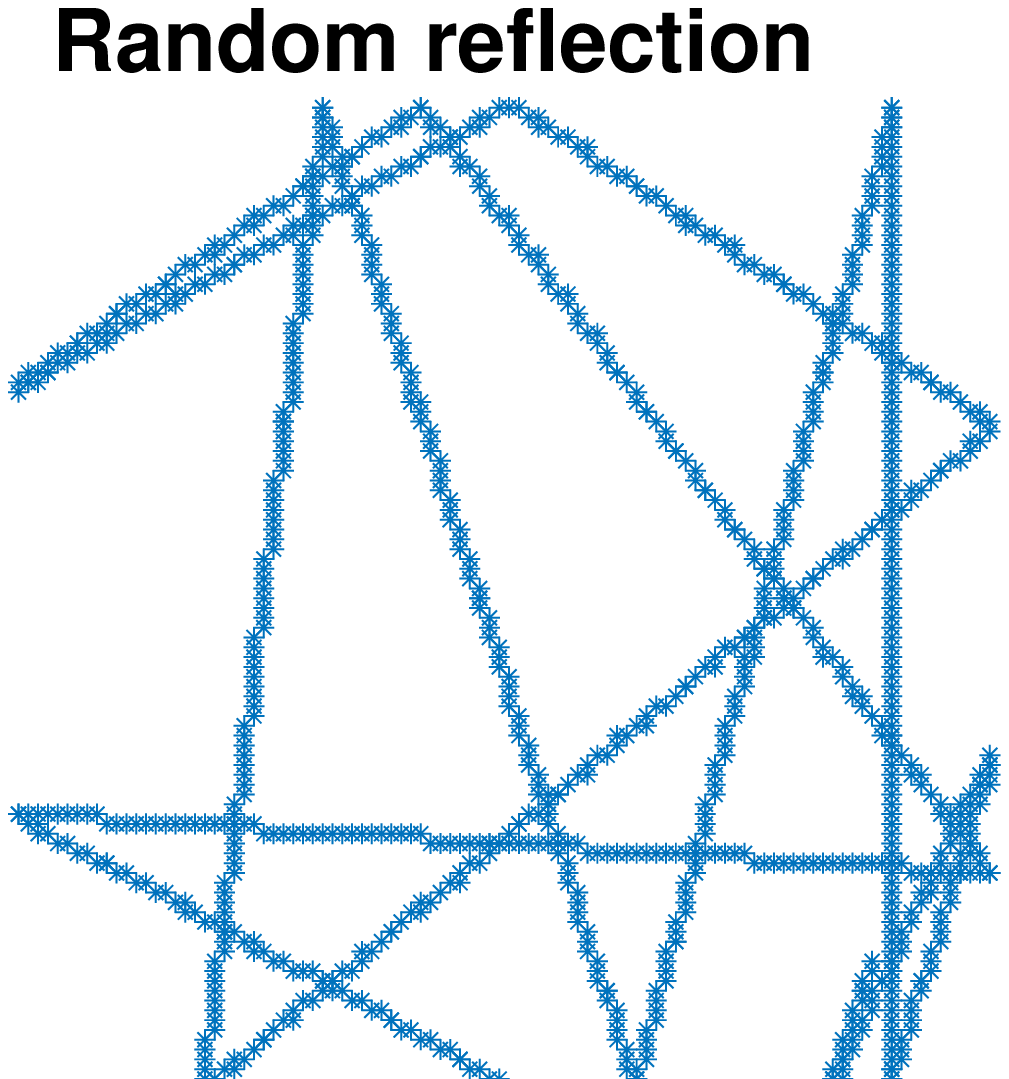}%
			\label{fig_third_case}}
		\quad
		\subfloat[]{\includegraphics[width=.12\textwidth]{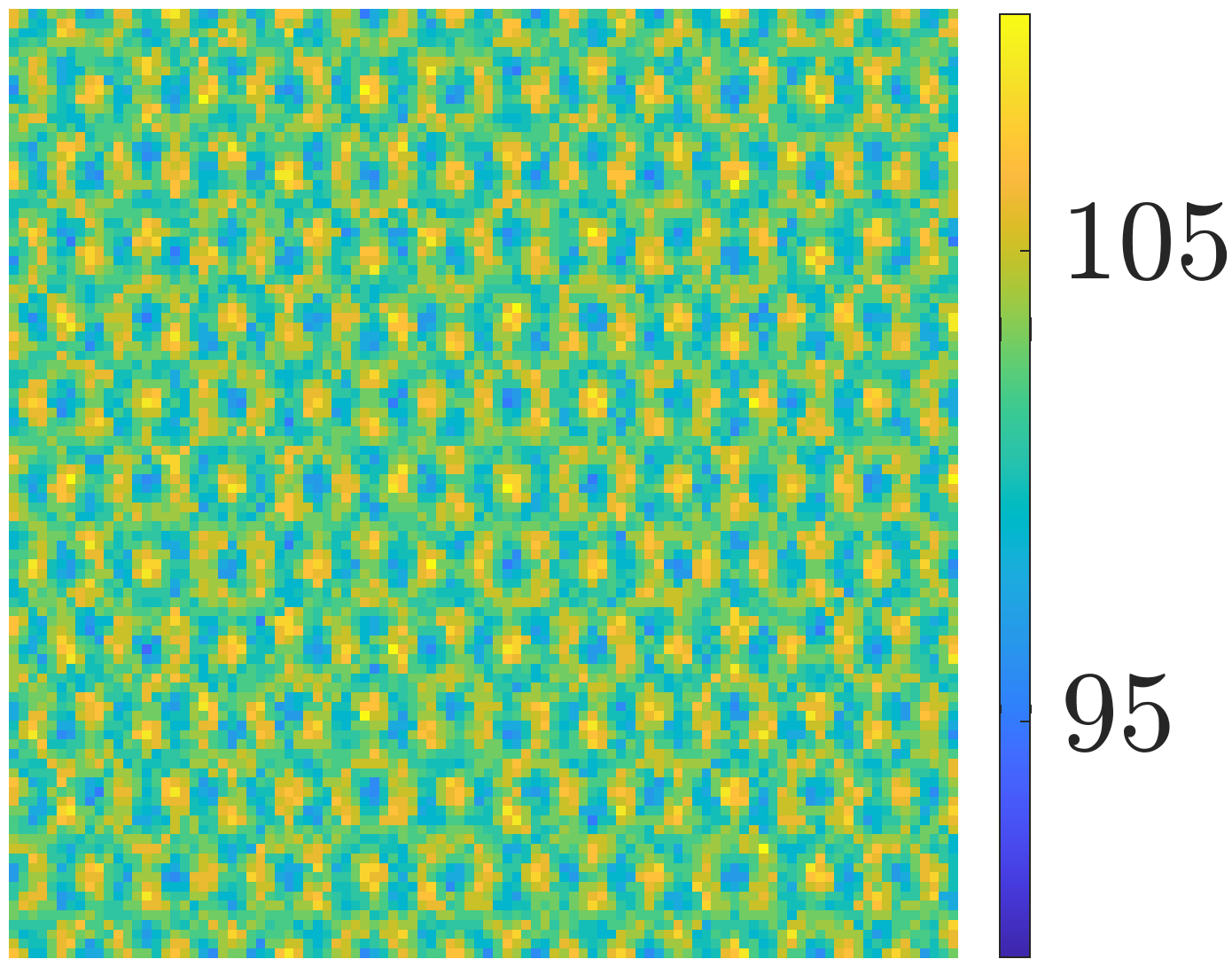}%
			\label{fig_fourth_case}}
		\quad
		\subfloat[]{\includegraphics[width=.12\textwidth]{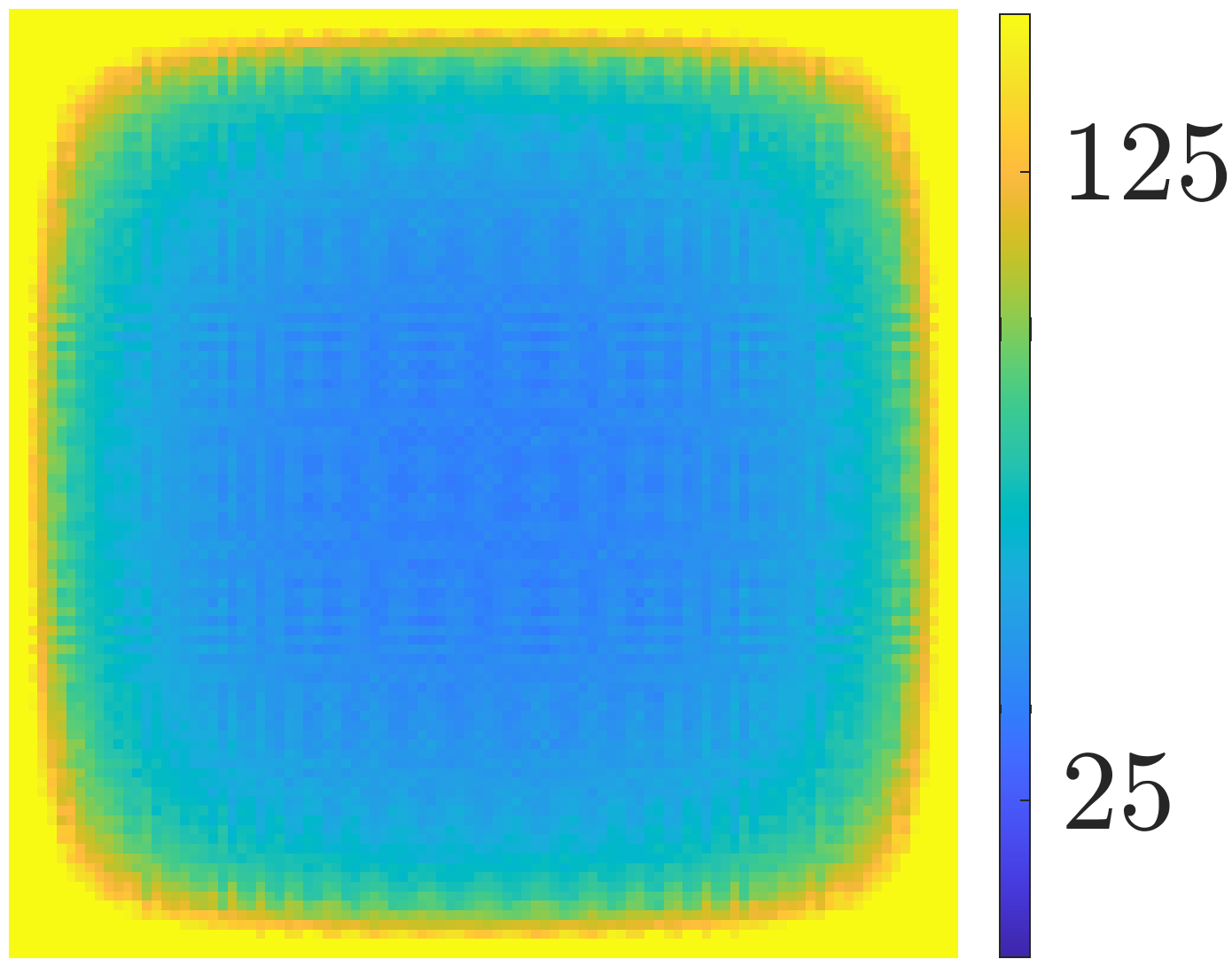}%
			\label{fig_fifth_case}}
		\quad
		\subfloat[]{\includegraphics[width=.12\textwidth]{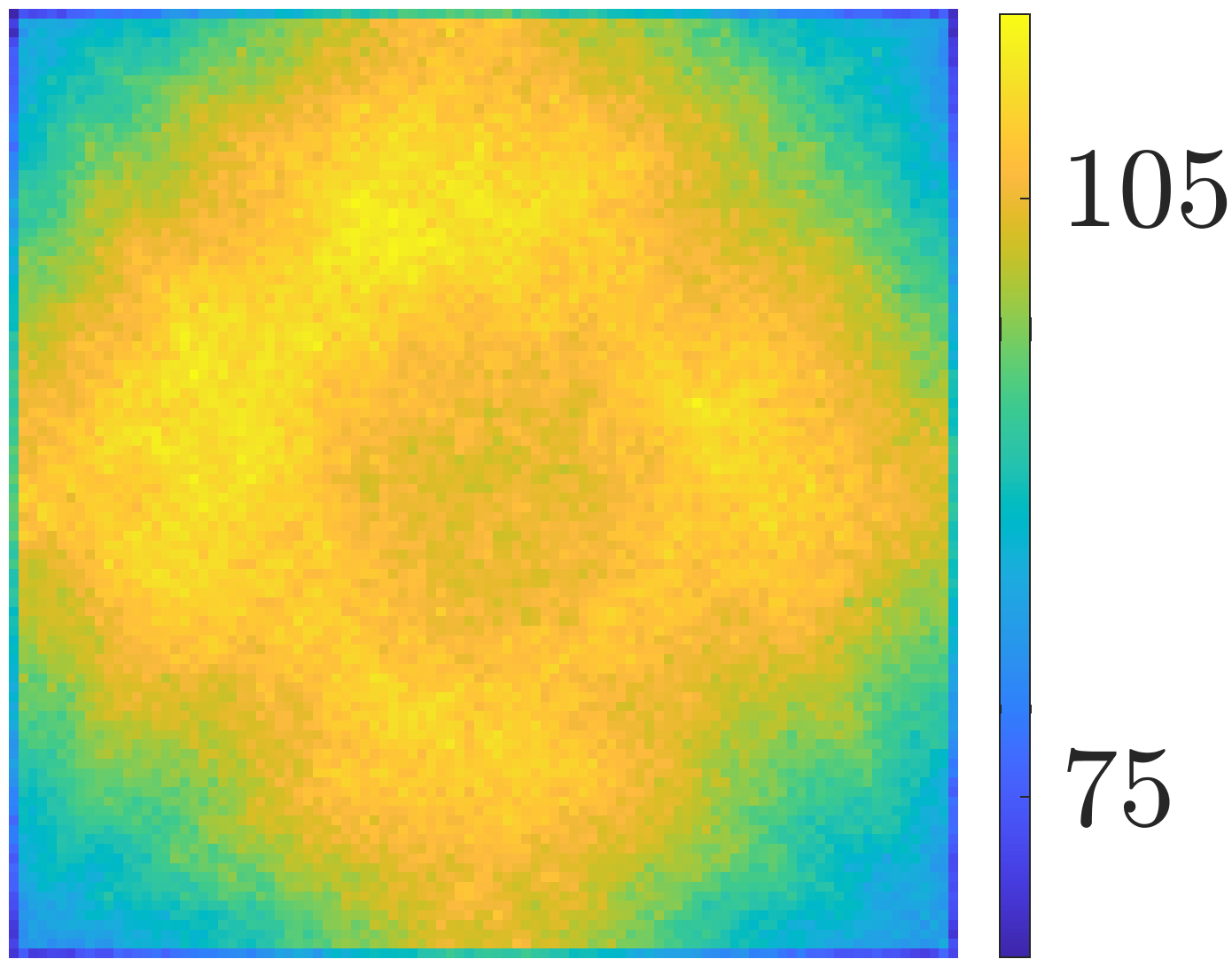}%
			\label{fig_sixth_case}}
		\caption{Visualization of different sub-sampling schemes. Top row: an xy-view of the first 1000 sampling points, bottom row: the sum of the sampled points along the z-axis for (a)+(d) triangle Lissajous (b)+(e) original Lissajous (c)+(f) random reflection. The colors in (d), (e) and (f) indicate the number of samples in the interferometer direction for each pixel in the $xy$-plane. Note that (e) has a different color scale than (d) and (f).}
		\label{Sampling}
	\end{figure}
	
	\subsubsection*{Random}
	
	According to \cite{candes2010matrix}, samples in a data matrix are chosen randomly to avoid information loss when recovering a low-rank matrix. The most obvious solution of sub-sampling is therefore a randomly selected subset from the full data set. We refer to \textit{random} if the samples were selected uniformly along the interferometer axis and within the image plane.
	
	\subsubsection*{Original Lissajous}
	
	The usage of random sub-sampling does not necessarily lead to significant savings in scan time. For randomly selected nano-FTIR measurements each sample would need to be approached by nanopositioning systems separately which is time-consuming. In addition, transit times between the samples are not used for measurement. We therefore consider routes along which the stages move, which usually occurs in a continuous manner. First, we consider an \textit{original Lissajous} sub-sampling. A 2D Lissajous curve \cite{tuma2012high} is the trajectory of a moving point whose coordinates are simple harmonic motions. The parametric equation is given by
	\begin{equation*}
	 t\mapsto \begin{pmatrix} A_0 + A\,\text{sin}(\omega_1t+\varphi_1) \\
	 	B_0 + B\,\text{sin}(\omega_2t+\varphi_2) \end{pmatrix}\,,
	\end{equation*}
	for $t\in[0,\infty)$. 2D Lissajous curves have been applied to several scanning techniques, e.g., in the fields of medical imaging \cite{feng2003single}, atomic force microscopy \cite{tuma2012high} and magnetic particle imaging \cite{werner2017first}. We will expand 2D Lissajous to 3D Lissajous by simply expanding the parametric equation as follows
	\begin{equation*}
		t\mapsto \begin{pmatrix} A_0 + A\,\text{sin}(\omega_1t+\varphi_1) \\
			B_0 + B\,\text{sin}(\omega_2t+\varphi_2)\\
		C_0 + C\,\text{sin}(\omega_3t+\varphi_3) \end{pmatrix}\,.
	\end{equation*}
	Figure~\ref{fig_second_case} shows an $xy$-view of the first 1000 sampling points along the 3D Lissajous curve. More details of the employed scheme can be found in the Appendix.
	
	\subsubsection*{Triangle Lissajous}
	
	Although the samples of original Lissajous scatter across the entire image stack, sinusoidal waves fail to create a uniform sampling pattern. In Figure~\ref{fig_fifth_case} the sum along the $z$-axis of all sample points is shown. Clearly, more samples are located at the border of the image which potentially could lead to an undesirable biased image reconstruction.
	In \cite{lin2018spectroscopic} a 3D \textit{triangle Lissajous} sub-sampling scheme has been introduced for  
	spectroscopic laser-scanning imaging. In Figure~\ref{fig_first_case} an $xy$-view of the first 1000 sampling points along the triangular Lissajous curve is shown. It can be seen (cf. Figure~\ref{fig_fourth_case}) that the triangular wave design yields a much more uniform sampling density than original Lissajous sub-sampling.
	
	Both original Lissajous and triangle Lissajous trajectories are sampled in equidistant time steps. For each time step the interferogram value of the nearest point on the $xyz$-grid is recorded.
	
	\subsubsection*{Random reflection}
	
	Similar to triangle Lissajous we consider a so-called \textit{random reflection} sub-sampling. From a given starting point the samples are acquired on a 3D line until one of the borders is reached. Then, specified by a random angle, the line is reflected and new samples can be measured until the next border is reached. This is repeated until the total number of samples is reached. In Figure~\ref{fig_third_case} an example of the $xy$-view of the first 1000 sampling points is shown and in Figure~\ref{fig_sixth_case} the sum along the $z$-axis of all sample points. Contrary to original Lissajous more samples are now in the center area. 
	
	\subsubsection*{White light}
	
	Finally, we want to introduce the so-called \textit{white light} sub-sampling. The idea here is to measure full $xy$-planes but only for some randomly chosen points along the interferometer axis. This means for some points along the $z$-axis no points will be measured in the image plane, while for others the entire $xy$-plane is recorded.
		
	\section{Results and Discussion}
	
	To test the influence of the different sub-sampling schemes, five subsets of the whole data set were created. The points belonging to each subset are chosen accordingly to one of the five sub-sampling schemes presented above using a sub-sampling rate of 10\%. 
	
	In the following, we will assess the quality of the low-rank matrix reconstructions in terms of the available full data set. Furthermore, we want to particularly compare the results when using a random sub-sampling scheme to the results for the other schemes.
	
\begin{figure*}[!t]
		\centering
		\includegraphics[width=.9\textwidth]{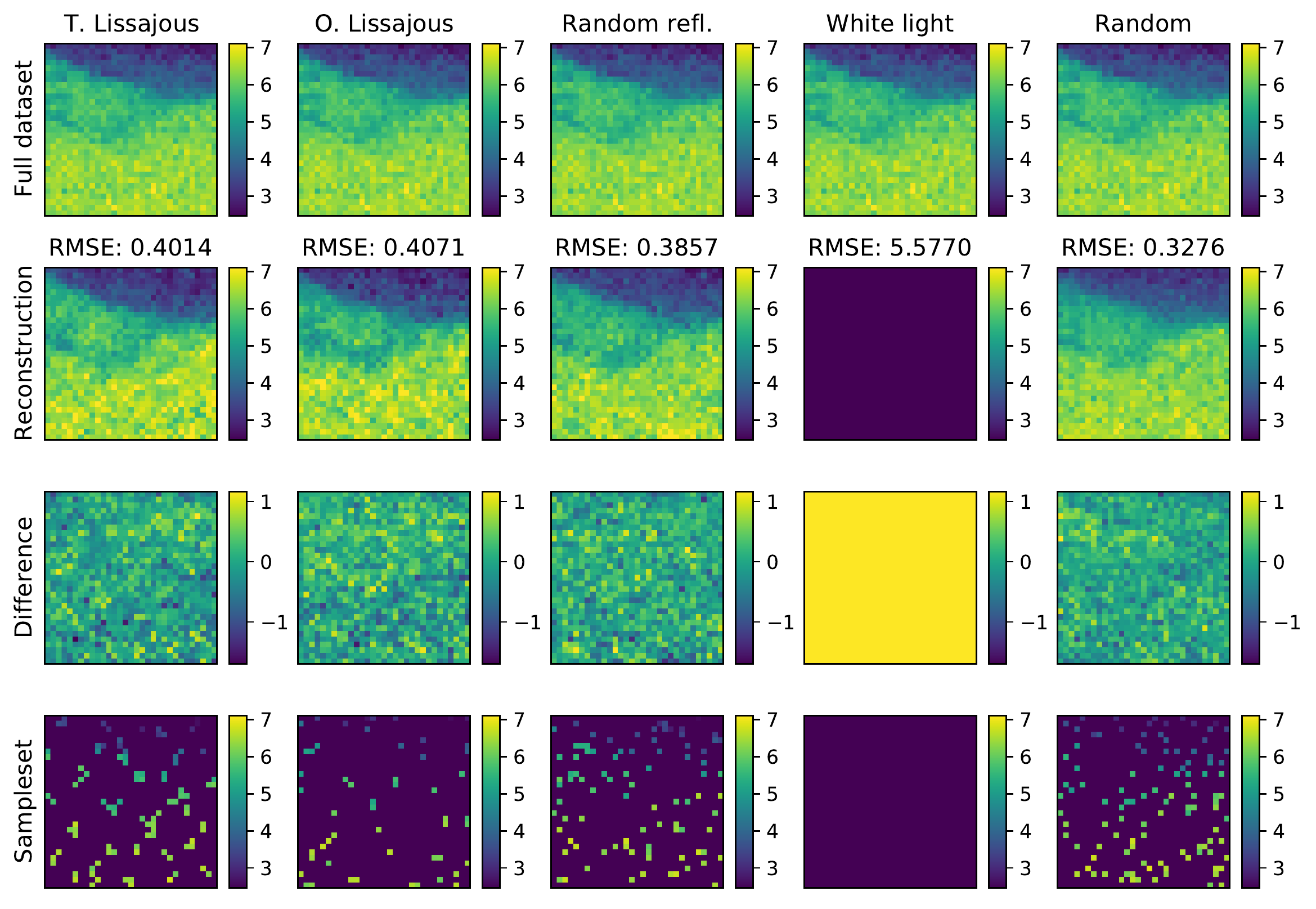}
		\caption{Reconstruction results for the different sub-sampling schemes at interferogram value of 1021.6 $\mu$m 
		using 10\% of the full data. First column: triangle Lissajous, second column: original Lissajous, third column: random reflection, fourth column: white light and fifth column: random sub-sampling. The number above the reconstructed data indicates the root-mean square error at the interferogram value of 1021.6 $\mu$m.}
		\label{Results_10}
	\end{figure*}

	In Figure \ref{Results_10} the $xy$-planes at the interferogram value of 1021.6 $\mu$m 
	are shown for the full data set, the low-rank reconstructions, the differences between them, and the (sub-sampled) data set for each sub-sampling scheme. 
	
	A summary of different similarity measures between the full data set and low-rank reconstructions can be found in Table~\ref{Table1}. The relative errors are calculated as $||\hat{X}-X||_F/||X||_F$, where $||\cdot||_F$ denotes the Frobenius norm, and where $\hat{X}$ and $X$ denote the reconstructed and the full data set, respectively. Note that the correlation coefficient cannot be calculated for white light sub-sampling since the reconstruction is 0 everywhere for any $z$-plane where no sample was measured.
	
	\begin{table}
		\begin{center}
			\caption{Similarity measures for the five different low-rank reconstructions. Each value represents the average over the respective measure for the reconstructed planes from 415.6 $\mu m$ to 1166.2 $\mu m$.
			}
			\label{Table1}
			\begin{tabular}{| c | c | c | c | c |}
				\hline
				Sub-sampling & Root-mean & 
				Relative & Correlation\\
				scheme & square error & error & coefficient \\
				\hline
				Triangle Lissajous& 0.30 & 0.48 
				& 0.68\\
				\hline
				Original Lissajous& 0.32 & 0.50 & 0.66 \\ 
				\hline
				Random reflection& 0.30 & 0.47 & 0.68 \\
				\hline
				White light& 1.18 & 0.93 & -\\
				\hline
				Random& 0.29 & 0.46 & 0.69\\
				\hline 
			\end{tabular}
		\end{center}
	\end{table}
	
	While the results of triangle Lissajous, original Lissajous and random reflection deliver similar results compared to a random sub-sampling, the results from the white light reconstruction are extremely poor.
	
	Note that Figure~\ref{Results_10} relates to an $xy$-plane that included no samples for the white light sub-sampling scheme. When looking at results of the white light sub-sampling for $xy$-planes that were fully measured, the reconstruction is almost perfect there.
	
\begin{figure}[!t]
		\centering
		\includegraphics[width=.4\textwidth]{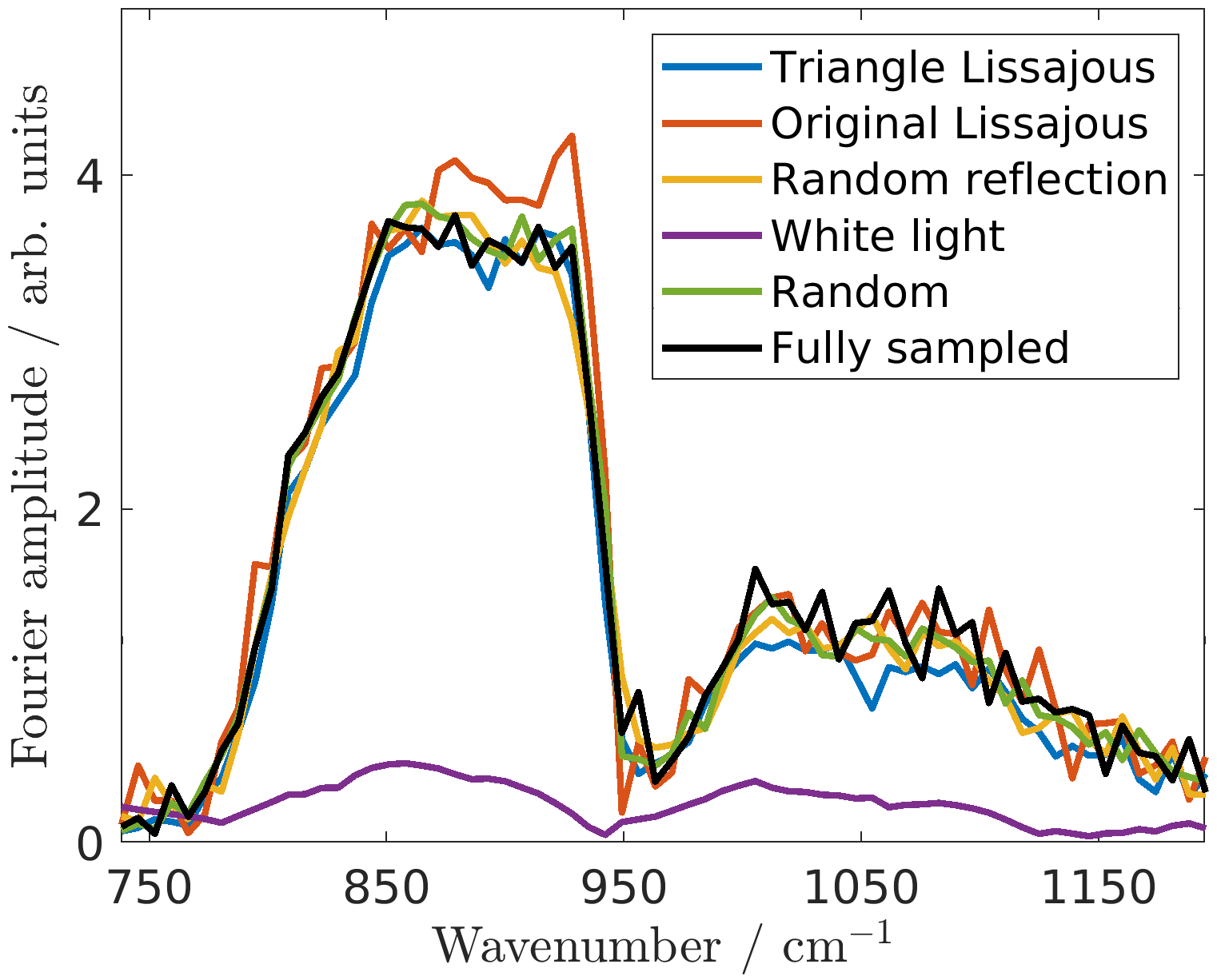}
		\caption{Reconstructed spectra from 731.1 cm$^{-1}$ to 1188.1 cm$^{-1}$ 
		at pixel (4,25) for the different sub-sampling schemes. Note that pixel (4,25) corresponds to point (a) in Figure~\ref{Results_spectra_peak}.}
		\label{Results_spectra_all}
	\end{figure}

	In Figure \ref{Results_spectra_all} the resulting spectra of the five reconstructions are shown for pixel (4, 25) and from 731.1 cm$^{-1}$ to 1188.1 cm$^{-1}$ 
	together with the spectrum of the full data set. Similar to the above results, the white light reconstruction is extremely poor whereas the other sub-sampling methods yield comparable results. Especially when comparing the random sub-sampling to triangle Lissajous and original Lissajous, the spectra are similar to each other.
		
\begin{figure}[!t]
		\centering
		\includegraphics[width=.45\textwidth]{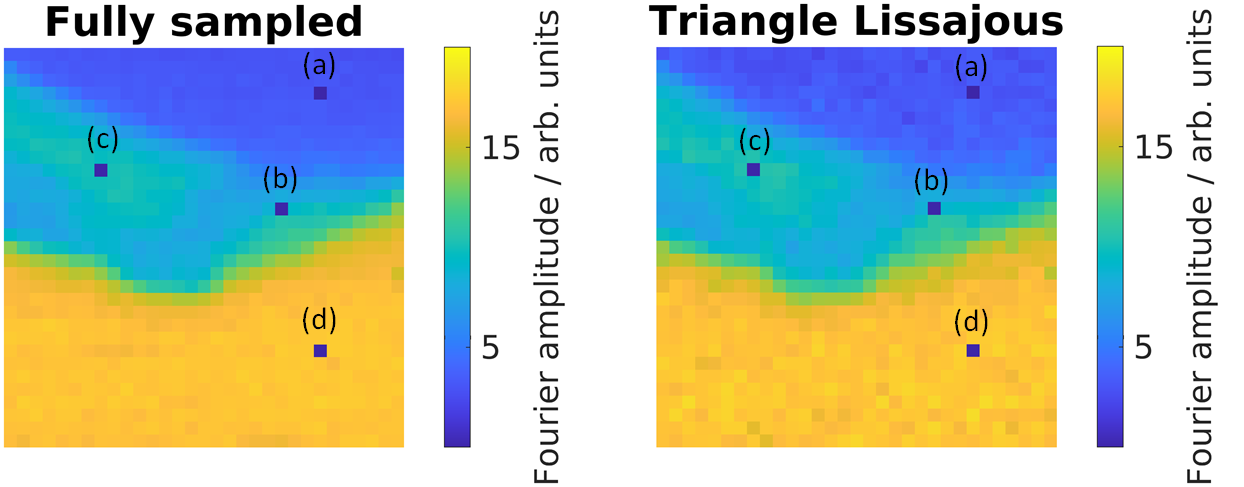}
		\caption{Amplitude of the spectra at peak position 920.9 cm$^{-1}$ for fully sampled data and sub-sampled data using a triangle Lissajous scheme. Four different pixel locations (a)-(d) are marked and evaluated in Figure~\ref{Results_spectra_ampl_phase}.}
		\label{Results_spectra_peak}
	\end{figure}

	In Figure~\ref{Results_spectra_peak} the amplitudes of the spectra at 920.9 cm$^{-1}$ for fully sampled data and triangle Lissajous sub-sampling are shown and appear to be in agreement.

\begin{figure*}[!t]
		\centering
		\includegraphics[width=.9\textwidth]{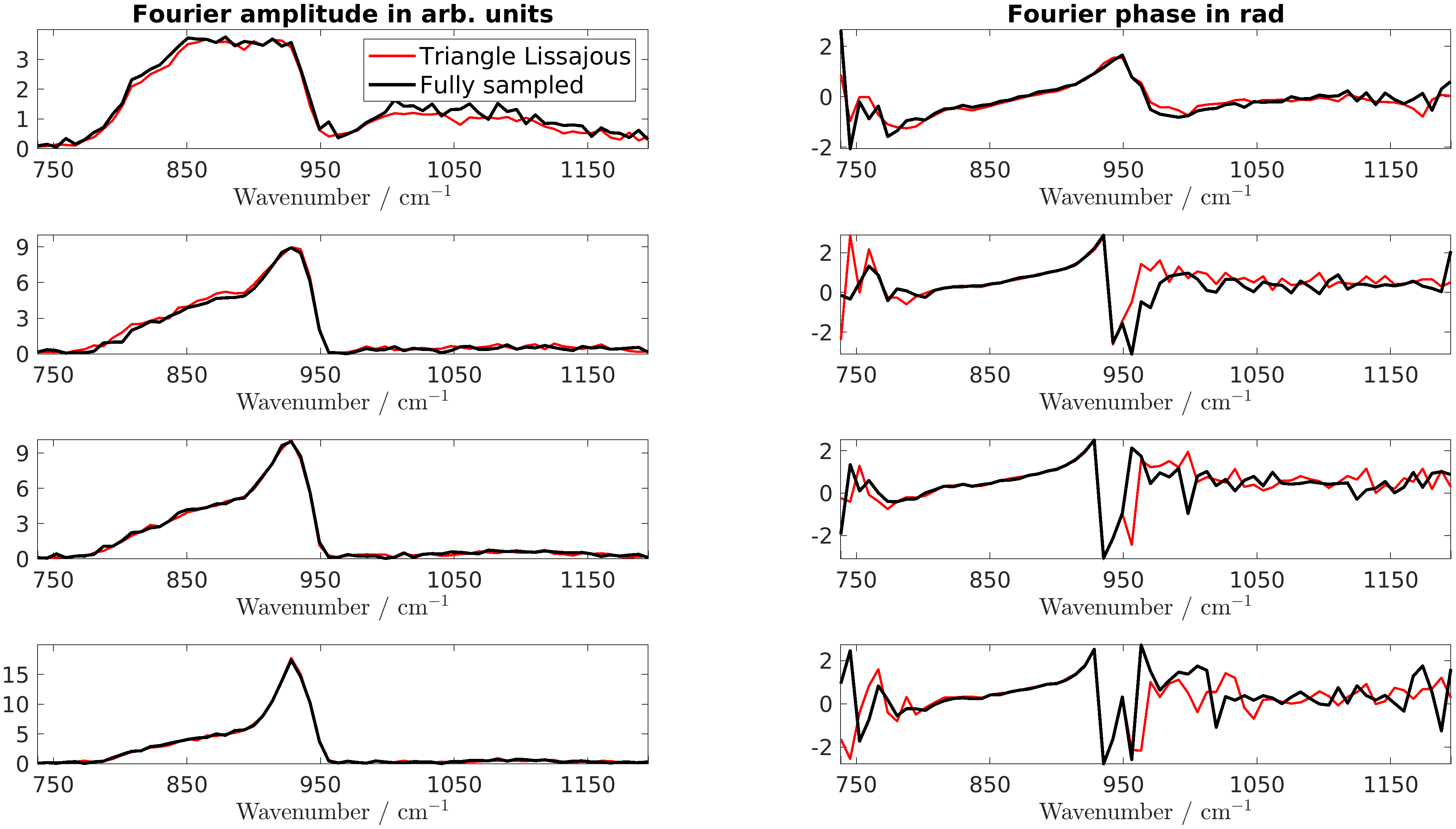}
		\caption{Amplitude and phase of the reconstructed spectra at different pixel locations for triangle Lissajous sub-sampling and fully sampled data. The pixel locations in the $xy$-plane are (a) (4,25), (b) (13,22), (c) (10,8), and (d) (24,25).}
		\label{Results_spectra_ampl_phase}
	\end{figure*}

	We choose four different points each characterizing a special part of the spectrum and compare the Fourier amplitude and the Fourier phase of the spectra from 731.1 cm$^{-1}$ to 1188.1 cm$^{-1}$
	(cf. Figure~\ref{Results_spectra_ampl_phase}). The quality of the low-rank reconstruction for the triangle Lissajous sub-sampling leads to an agreement to the fully sampled data set.

	In order to verify that the results for sampling schemes with random elements (i.e. random and random reflection sub-sampling) are representative, two further, randomly drawn subsets were created for each scheme and used for the reconstruction of the full data set. It turned out that the above results do not change significantly. In particular, we obtained the same numbers as in Table~\ref{Table1}.
	
\begin{figure*}[!t]
		\centering
		\includegraphics[width=.9\textwidth]{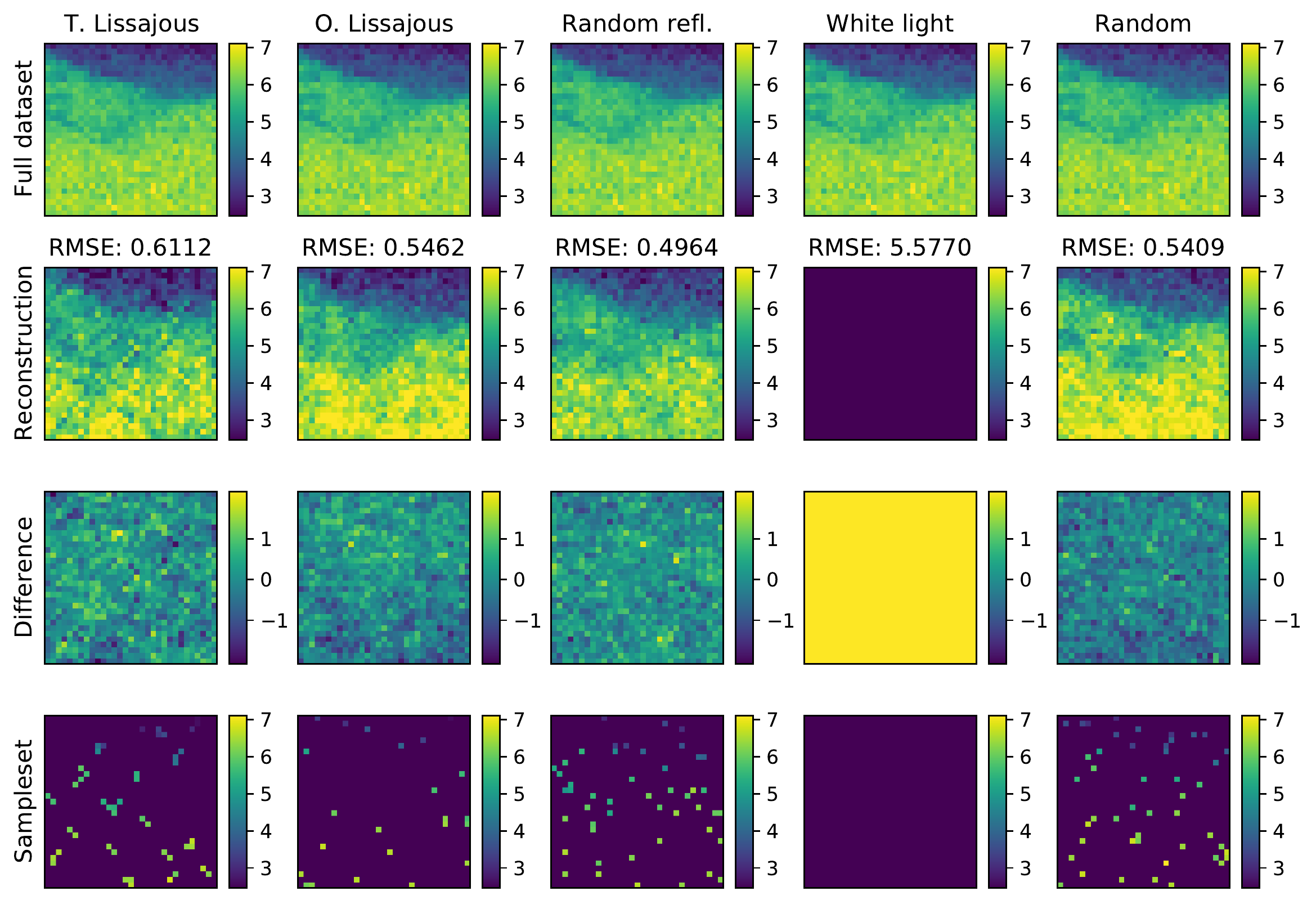}
		\caption{Reconstruction results for the different sub-sampling schemes at the interferogram value of 1021.6 $\mu$m 
		using 5\% of the full data. First column: triangle Lissajous, second column: original Lissajous, third column: random reflection, fourth column: white light and fifth column: random sub-sampling.}
		\label{Results_5}
	\end{figure*}

	We also created five data sets with a sub-sampling rate of 5\%. Results of the low-rank reconstructions can be seen in Figure~\ref{Results_5}. As expected, reducing the amount of data leads to a deterioration of the reconstruction quality. This is also indicated by the increased root-mean square errors of the reconstructions. Nevertheless, for triangle Lissajous, original Lissajous, and random reflection the main features are still clearly visible and comparable to a random sub-sampling. The white light sub-sampling is still delivering poor results.
		
	\section{Conclusion \& Outlook}
	
	Different sub sampling schemes have been compared for compressive nano-FTIR measurement. Triangle Lissajous, original Lissajous, and random reflection have been shown to lead to a suitable data reconstruction for a sub-sampling rate of 10\%. Even with a  sub-sampling rate of 5\% reasonable results could be achieved. All three sub-sampling schemes yield similar results. Furthermore, these results are in agreement with results of random sub-sampling. We therefore have presented three practically relevant sub-sampling schemes which could lead to tremendous time savings for nano-FTIR measurements. 
	
	Future work could apply the sub-sampling schemes to other nano-FTIR measurements and even other scanning probe techniques. Additionally, a hardware implementation of these schemes is necessary to compare the results regarding their acquisition times. We expect that due to the absence of idle times and the possibility of continuous stage movement the reduction in acquisition times approaches the sub-sampling rate.
	
	\section*{Acknowledgments}
	We gratefully acknowledge fruitful discussions with Alexander Govyadinov and Markus Raschke. We acknowledge financial support by Deutsche Forschungsgemeinschaft (Grants EL 492/1-1, RU 420/13-1). Bernd Kästner has received funding within the project 20IND09 PowerElec from the EMPIR programme co-financed by the Participating States and from the European Union's Horizon 2020 research and innovation programme.
	\bibliographystyle{ieeetr} 
	\bibliography{Library}
	
	\section*{Appendix}
	\subsection*{Details about the sample}
	Epitaxial graphene was grown on the Si-face of SiC samples (5 × 10) mm\textsuperscript{2} cut from a semi-insulating 6H-SiC wafer with a nominal miscut angle of about 0.06$^\circ$ toward the (1100) plane. The graphene samples were prepared using the polymer-assisted sublimation growth (PASG) technique, which involves polymer adsorbates formed on the SiC surface by liquid phase deposition from a solution of a photoresist (AZ5214E) in isopropanol followed by sonication and short rinsing with isopropanol. The graphene layer growth was processed at 1750 $^\circ$C (argon atmosphere $\approx$1 bar, 6 min, zero argon flow) with pre-vacuum-annealing at 900 $^\circ$C~\cite{Kruskopf2016, Pakdehi2019}. The applied PASG method allows for the growth of large-area homogeneous monolayer graphene, with almost an isotropic resistance characteristics~\cite{Pakdehi2018}. The fabrication of a large-area two-dimensional gallium layer at the interface of  SiC and graphene has been realized by using the liquid metal intercalation technique (LiMIT)~\cite{wundrack2021liquid}. This method achieves the lateral intercalation and diffusion of Ga atoms at room temperature resulting in the conversion of epitaxial graphene to quasi-freestanding bilayer graphene (QFBLG). The average thickness of the confined gallium layer after the intercalation process is approx. 1 nm at the interface between SiC and graphene. In addition, Van der Pauw measurements revealed low carrier mobility as well as strong electron doping of n $\approx$ 4.5$\times 10 \textsuperscript{12} $cm$\textsuperscript{-2} $ in the SiC- confined Ga – QFBLG sample (SiC/2DGa/QFBLG) at room temperature~\cite{wundrack2021liquid}.
	
	\subsection*{Details about the employed original Lissajous sub-sampling scheme}
		In Section 2D a general description of the original Lissajous sub-sampling scheme is given. In the present case for a data cube of dimension $31\times 31\times 1024$ the following parameters have been used for the employed original Lissajous scheme: $A_0=15.5$, $B_0=15.5$, $C_0=512$, $A=97.3894$, $B=97.3894$, $C=3216.9908$, $\omega_1=0.0119$, $\omega_2=0.0084$, $\omega_3=0.0002$, $\varphi_1=0.75$, $\varphi_2=0.75$, and $\varphi_3=0.75$.
		
	
		
	\vfill
	
\end{document}